\documentclass[aps,prx,twocolumn,superscriptaddress,showpacs,showkeys,
notitlepage]{revtex4-1}

\usepackage{graphicx}
\usepackage{dcolumn}
\usepackage{bm}
\usepackage{float}
\usepackage{amsmath}
\let\oldAA\AA
\renewcommand{\AA}{\text{\normalfont\oldAA}}
\usepackage{xcolor}
\usepackage{textcomp}
\usepackage{textgreek}
\usepackage[symbol]{footmisc}

\begin{document}

\title{Pinning of Helimagnetic Phase Transitions in Zn-Substituted Skyrmion Host Cu$_2$OSeO$_3$}

\author{M. T. Birch}
\address{Durham University, Centre for Materials Physics, Durham, DH1 3LE, United Kingdom}
\address{Diamond Light Source, Didcot, OX11 0DE, United Kingdom}

\author{S. H. Moody}
\address{Durham University, Centre for Materials Physics, Durham, DH1 3LE, United Kingdom}

\author{M. N. Wilson}
\address{Durham University, Centre for Materials Physics, Durham, DH1 3LE, United Kingdom}

\author{M. Crisanti}
\address{Department of Physics, University of Warwick, Coventry, CV4 7AL, United Kingdom}
\address{Institut Laue-Langevin, CS 20156, 38042 Grenoble Cedex 9, France}

\author{O. Bewley}
\address{Durham University, Centre for Materials Physics, Durham, DH1 3LE, United Kingdom}

\author{A. \v{S}tefan\v{c}i\v{c}}\thanks{Current Address: Electrochemistry Laboratory, Paul Scherrer Institut, CH-5232 Villigen PSI, Switzerland}
\address{Department of Physics, University of Warwick, Coventry, CV4 7AL, United Kingdom}

\author{G. Balakrishnan}
\address{Department of Physics, University of Warwick, Coventry, CV4 7AL, United Kingdom}

\author{R. Fan}
\address{Diamond Light Source, Didcot, OX11 0DE, United Kingdom}

\author{P. Steadman}
\address{Diamond Light Source, Didcot, OX11 0DE, United Kingdom}

\author{D. Alba Venero}
\address{ISIS Neutron and Muon Source, Rutherford Appleton Laboratory, Didcot, OX11 0QX, UK.}

\author{R. Cubitt}
\address{Institut Laue-Langevin, CS 20156, 38042 Grenoble Cedex 9, France}

\author{P. D. Hatton}
\address{Durham University, Centre for Materials Physics, Durham, DH1 3LE, United Kingdom}

\begin{abstract}
Magnetic skyrmions are nano-sized topological spin textures stabilized by a delicate balance of magnetic energy terms. The chemical substitution of the underlying crystal structure of skyrmion-hosting materials offers a route to manipulate these energy contributions, but also introduces additional effects such as disorder and pinning. While the effects of doping and disorder have been well studied in B20 metallic materials such as Fe$_{1-x}$Co$_x$Si and Mn$_{1-x}$Fe$_x$Si, the consequences of chemical substitution in the magnetoelectric insulator Cu$_2$OSeO$_3$ have not been fully explored. In this work, we utilize a combination of AC magnetometry and small angle neutron scattering to investigate the magnetic phase transition dynamics in pristine and Zn-substituted Cu$_2$OSeO$_3$. The results demonstrate that the first order helical-conical phase transition exhibits two thermally separated behavioural regimes: at high temperatures, the helimagnetic domains transform by large-scale, continuous rotations, while at low temperatures, the two phases coexist. Remarkably, the effects of pinning in the substituted sample are less prevalent at low temperatures, compared to high temperatures, despite the reduction of available thermal activation energy. We attribute this behaviour to the large, temperature-dependent, cubic anisotropy unique to Cu$_2$OSeO$_3$, which becomes strong enough to overcome the pinning energy at low temperatures. Consideration and further exploration of these effects will be crucial when engineering skyrmion materials towards future applications.
\end{abstract}

\maketitle

\section{Introduction}
Magnetic skyrmions are nanoscale, topologically protected particles \cite{rosler_spontaneous_2006}. In conductive samples, it has been shown that under the application of a current, they have high mobility \cite{iwasaki_current-induced_2013}. These properties have made skyrmions the focus of recent research due to their potential application in future spintronic devices \cite{nagaosa_topological_2013}. The presence of the Dzyaloshinskii-Moriya Interaction (DMI) is vital for the formation magnetic skyrmions in a range of material systems \cite{Dzyaloshinskii}. Interfacial DMI, induced by spin-orbit coupling at the interface between layers in thin films, produces N\'eel-type skyrmions \cite{heinze_spontaneous_2011,fert_skyrmions_2013,woo_observation_2016}. In chiral magnets, such as MnSi \cite{muhlbauer_skyrmion_2009,neubauer_topological_2009}, MnGe \cite{tanigaki_realspace_2015}, Fe$_{1-x}$Co$_x$i \cite{munzer_skyrmion_2010,yu_real-space_2010}, FeGe \cite{wilhelm_precursor_2011,yu_near_2011}, the multiferroic Cu$_2$OSeO$_3$ \cite{seki_observation_2012} and the $\beta$-Mn type Co-Zn-Mn alloys \cite{tokunaga_new_2015}, the lack of centrosymmetry in the underlying atomic lattice gives rise to bulk DMI, leading to the formation of Bloch-type skyrmions, or in the case of the polar materials GaV$_4$Se$_8$ and GaV$_4$S$_8$, N\'eel-type skyrmions \cite{kezsmarki_neel_2015,ruff_multiferroicity_2015}. 

In these bulk materials, the interplay of the DMI with the exchange and dipolar interactions, the Zeeman energy, and the magneto-crystalline anisotropies, is responsible for a range of magnetic structures and phenomena \cite{nagaosa_topological_2013}. At zero applied magnetic field the helical state is realized by the competition of the exchange and DMI \cite{bak_theory_1980}. Typically, multiple degenerate helical domains order and orient to point their spins along easy planes, as determined by the underlying cubic anisotropy. Under an applied magnetic field, the Zeeman energy lifts the helical domain degeneracy, reorienting the magnetic structures such that their spin propagation vectors point along the field direction, forming the conical state. Close to the Curie temperature, $T_c$, critical thermal fluctuations are responsible for stabilizing the hexagonal skyrmion lattice against the conical state in a small range of applied magnetic field \cite{nagaosa_topological_2013}. However, it has been demonstrated that by cooling the system under an applied magnetic field, skyrmions may survive in a metastable state beyond the typical equilibrium range \cite{karube_robust_2016,karube_skyrmion_2017,oike_interplay_2016,kagawa_current-induced_2017,milde_unwinding_2013,okamura_transition_2016}.

The role of cubic anisotropy has been highlighted in a number of recent studies. By utilizing the wide temperature and applied field range of the metastable skyrmion state, a transition from the typical hexagonal arrangement to a square skyrmion lattice has been observed in both MnSi \cite{nakajima_skyrmion_2017}, and Co$_8$Zn$_8$Mn$_4$ \cite{karube_robust_2016}, induced by the increased effective anisotropy strength at low temperatures. Meanwhile, in Cu$_2$OSeO$_3$, the strong spin-orbit coupling results in a comparatively large, temperature-dependent cubic anisotropy \cite{halder_thermodynamic_2018}. At low temperatures, this anisotropy is of sufficient strength to rotate the conical state away from the applied magnetic field direction, forming the tilted conical state \cite{qian_new_2018}, and is further responsible for the stabilization of the equilibrium low temperature skyrmion state over much of the temperature-field phase diagram \cite{chacon_observation_2018,bannenberg_multiple_2019}. These discoveries highlight the rich variety of spin textures accessible by manipulating the balance of interaction energies in chiral magnets. 

One way to control the relative contribution of these energy terms is by varying the elemental composition of skyrmion-hosting materials via chemical substitution or doping. Compositional effects have been studied in a range systems, such as Fe$_{1-x}$Co$_x$Si \cite{bauer_history_2016,bannenberg_extended_2016,bannenberg_magnetic_2016}, Mn$_{1-x}$(Fe/Co)$_x$Si \cite{grigoriev_helical_2009,franz_realspace_2014,bannenberg_magnetization_2018,bannenberg_evolution_2018}, Mn$_{1-x}$Fe$_x$Ge \cite{grigoriev_chiral_2013,shibata_towards_2013,koretsune_control_2015,grigoriev_spinwave_2018}, MnSi$_{1-x}$Ge$_x$ \cite{fujishiro_topological_2019},  Co$_{10-x}$Zn$_{10-y}$Mn$_{x+y}$ \cite{tokunaga_new_2015, karube_skyrmion_2017, takagi_spinwave_2017}, Co$_{8-x}$(Fe/Ni/Ru)$_x$Zn$_8$Mn$_4$ \cite{karube_controlling_2018}, GaV$_4$(S$_{8-x}$Se$_x$) \cite{franke_magnetic_2018,hicken_magnetizm_2020}, [Cu$_{1-x}$(Zn/Ni)$_x$]$_2$OSeO$_3$ \cite{wu_physical_2015,chandrasekhar_effects_2016,stefancic_origin_2018,birch_increased_2019,wilson_measuring_2019,sukhanov_increasing_2019} and Cu$_2$OSe$_{1-x}$Te$_x$O$_3$ \cite{han_scaling_2020}. The properties altered include: variation of $T_c$, indicating a change in the exchange interaction strength; alteration of the magnitude and sign of the DMI, leading to changes in the helical and skyrmion lattice periodicity and chirality; switching of the magnetic easy-axes, and therefore the helical ground state domain orientation, by tuning of the cubic anisotropy constants; and modification of the spin wave propagation. 

Furthermore, it was demonstrated that the disorder present in the underlying atomic lattice of these doped systems introduced pinning effects which allowed the formation of larger populations of metastable skyrmions when field cooling at lower cooling rates \cite{munzer_skyrmion_2010}, and increased both the annihilation and formation time of the skyrmion state \cite{birch_increased_2019,wilson_measuring_2019}. This pinning can affect other dynamic behaviour such as current-induced skyrmion motion - a phenomenon crucial to proposed skyrmion racetrack devices \cite{tomasello_strategy_2015, zhang_skyrmion_2015}. Extrinsic defects and impurities can attract or repel skyrmions \cite{fernandes_universality_2018}, or act as nucleation or annihilation sites \cite{fallon_controlled_2020}, which might be exploited for technological implementation. In combination with alterations to the energy balance mentioned above, these dynamical pinning effects must be considered when engineering the properties of skyrmion materials for future applications.

In this work, we utilize detailed AC magnetometry and small angle neutron scattering (SANS) to investigate the effect of doping, and the role of cubic anisotropy, in helimagnetic phase transitions in (Cu$_{1-x}$Zn$_x$)$_2$OSeO$_3$. Beginning with an explanation of the experimental methods in Sec. \ref{s:methods}, we proceed to detail the spin textures exhibited by Cu$_2$OSeO$_3$ in Sec. \ref{s:chiral}. Typical phase diagrams are considered in Sec. \ref{s:acsus}, revealing the AC susceptibility signals exhibited by each magnetic state. This is followed by Sec. \ref{s:tempdep}, where the detailed features of the real and imaginary components of the AC susceptibility measurements are examined and interpreted. Comparison of the helical-conical phase boundary in both samples reveal significant history-dependent pinning behaviour in the substituted sample. Further investigation of the dissipative effects indicated by peaks in the imaginary component reveals unexpected low temperature behaviour in both samples. 

The results of the SANS measurements are considered in Sec. \ref{s:sans}, with reference to the associated features of the AC susceptibility data in Sec. \ref{s:tempdep}. The volume fractions of all magnetic structures are assessed, revealing that the helical-conical phase transition exhibits two distinct, thermally-separated behavioural regimes. In Sec. \ref{s:cubicaniso}, we examine temperature dependent SANS measurements, and with comparison to previous studies of other chemically substituted skyrmion materials, discuss the observed behaviour in the context of the large, temperature-dependent cubic anisotropy exhibited by Cu$_2$OSeO$_3$, which may explain the unique behaviour observed in this study. Finally, we summarise the results and conclusions in Sec. \ref{s:conclusions}.

\section{Methods}
\label{s:methods}
Single crystals of Cu$_2$OSeO$_3$ were grown from polycrystalline powders using the chemical vapour transport method \cite{stefancic_origin_2018}. The elemental composition of the resulting crystals was determined by inductively-coupled plasma mass spectroscopy, as reported in previous work \cite{stefancic_origin_2018}. The samples chosen for this study were a pristine Cu$_2$OSeO$_3$ crystal, with $x=0.00$, or 0\% Zn-substitution, and a (Cu$_{1-x}$Zn$_x$)$_2$OSeO$_3$ crystal with $x=0.02$, or 2\% Zn-substitution.

Magnetometry measurements were performed with a superconducting quantum interference device magnetometer (MPMS3, Quantum Design), fitted with the AC magnetometry option. Temperature control was achieved by utilizing the inbuilt helium cryostat. The samples were attached to a quartz rod using GE varnish and mounted in the system with the $[110]$ crystal direction parallel to the applied magnetic field. Measurements of magnetization vs temperature determined the $T_\text{C}$ of the 0\% and 2\% samples to be 58.8~K and and 57.3~K respectively. AC susceptibility measurements were performed with an oscillating magnetic field amplitude of 0.1~mT at a frequency of 10~Hz, and all cooling procedures were performed at a rate of 40 K/min. 

\begin{figure}
\centering
\includegraphics[width=0.49\textwidth]{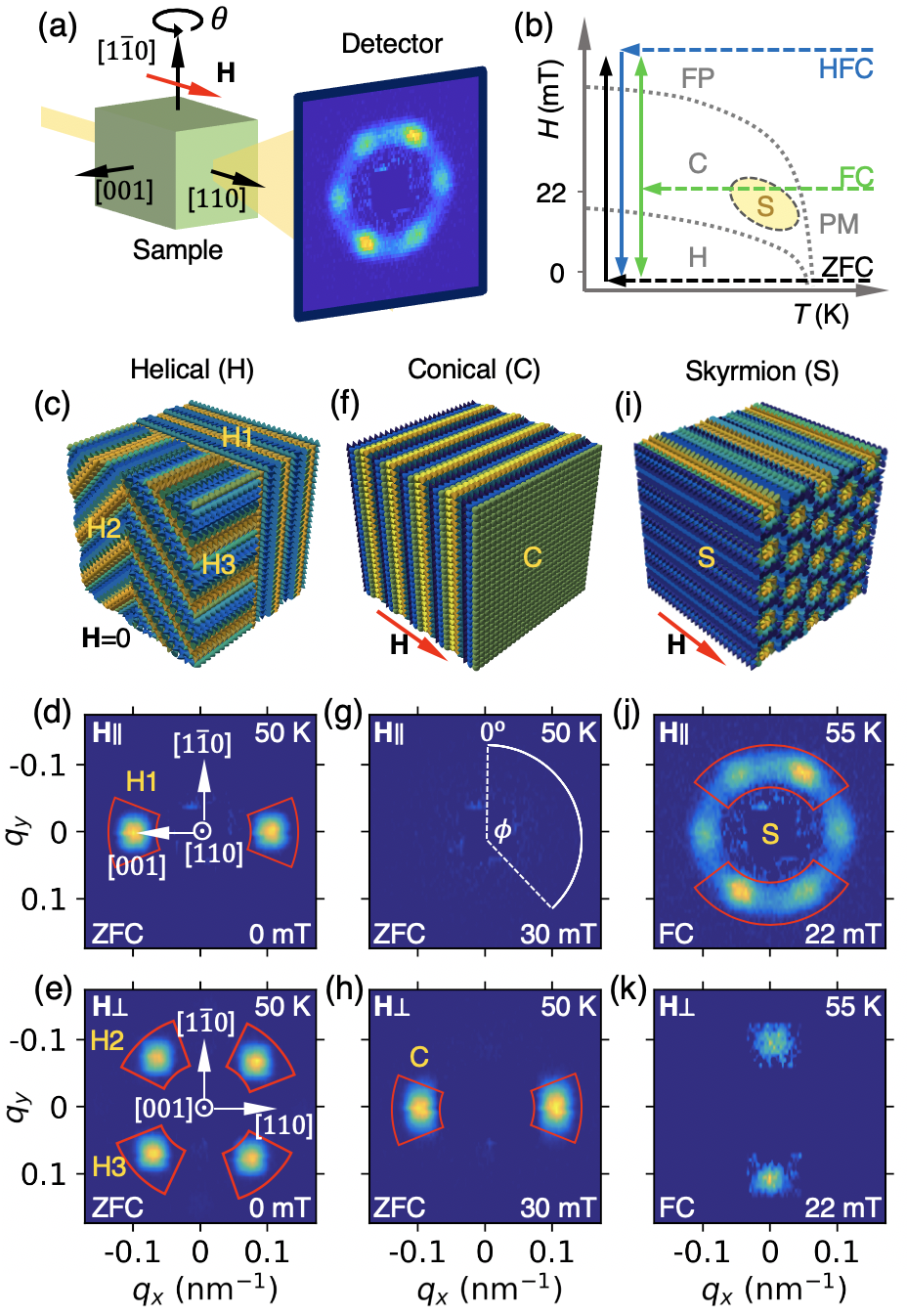}
\caption{(a), Schematic illustration of the small angle neutron scattering experiment setup. The sample and applied field can be rotated together through $\theta$ to measure with the magnetic field directed both parallel and perpendicular to the neutron beam. The relative orientation of the Cu$_2$OSeO$_3$ sample is shown. (b), An example, skyrmion material phase diagram, showing the helical (H), conical (C),skyrmion (S), uniform magnetization (UM) and paramagnetic (PM) states. Representations of the zero-field cooled (ZFC), high-field cooled (HFC) and field cooled (FC) procedures are shown. Three-dimensional visualizations of the spin textures and characteristic SANS patterns measured for the helical (c-e), conical (f-h) and skyrmion states (i-k) states, with the field both parallel (d,g,j) and perpendicular (e,h,k) to the neutron beam. Sector boxes indicate regions of scattering summed to report intensities in subsequent figures. The azimuthal angle around each SANS pattern, $\phi$, is illustrated in panel (g).}
\label{fig1}
\end{figure}

SANS measurements were performed on the D33 instrument at the Institut Laue-Langevin (ILL) reactor, and the ZOOM instrument \cite{ISIS_data} at the ISIS spallation source. The samples were attached to a 200 \textmu m thick aluminium plate and placed inside a helium cryostat equipped with superconducting magnets. At ILL, a neutron wavelength of 6.0~$\AA$ was selected, with a FWHM spread of $\lambda/\lambda=10\%$. Due to ISIS being a time-of-flight source, a range of neutron wavelengths were accepted between 1.75~$\AA$ and 17.5~$\AA$. Cooling procedures were formed at a rate of 7~K/min and 0.5~K/min at ILL and ISIS respectively. A schematic of the experimental setup is shown in Fig. \ref{fig1}(a). During diffraction measurements, coherent scattering is observed when the orientation of the magnetic structures fulfil the Bragg condition. For SANS, this occurs when the direction of periodicity of the magnetic structure is close to perpendicular to the incident neutron beam, resulting in detection of a diffraction peak, or magnetic satellite. 

Each Cu$_2$OSeO$_3$ crystal was positioned with the $[110]$ axis parallel to the applied magnetic field, and the ${[1\bar{1}0]}$ direction aligned with the vertical rotation axis, as shown in Fig. \ref{fig1}. A rotation around this axis through $\theta$ rotates both the sample and applied magnetic field simultaneously. All SANS patterns shown are the result of summing scans measured as a function of this rocking angle $\theta$. In this particular sample and field setup, scattering peaks from all ordered chiral magnetic structures exhibited by the sample may be observed by rotating the sample and field through 90 degrees, such that the applied field is positioned either parallel or perpendicular to the neutron beam. For the rest of the article, these two sample orientations will be referred to as the `field-parallel' and `field-perpendicular' configurations respectively.

In this study, we employed three distinct field-temperature protocols, as illustrated in Fig. \ref{fig1}(b). In the zero field-cooled (ZFC), field-cooled (FC) and high field-cooled (HFC) procedures, the sample was cooled to the target temperature under an applied magnetic field of 0, 20 and 200 mT respectively. After initializing the magnetic state under these conditions, measurements were carried out as a function of increasing (ZFC) and decreasing (HFC) applied magnetic field. In the case of the FC procedure, measurements were performed for increasing and decreasing magnetic field after two separate cooling procedures. Metastable skyrmions were formed only during the FC protocol, as this field-temperature cooling path took the sample through the equilibrium skyrmion phase.

\section{Chiral Magnetic States}
\label{s:chiral}

Illustrations of the spin textures exhibited by Cu$_2$OSeO$_3$, and their corresponding SANS patterns in the two experimental configurations, are displayed in Fig. \ref{fig1}. At zero magnetic field, the helical structure is the equilibrium state, consisting of a continuous rotation of spins orthogonal to a propagation vector. In Cu$_2$OSeO$_3$, this vector aligns along the $\langle100\rangle$ crystal axes due to the cubic anisotropy present in the system, giving rise to three distinct helical domains, labelled H1, H2 and H3 in Fig. \ref{fig1}(c). Scattering from these domains can be observed in the field-parallel configuration (H1), or in the field-perpendicular configuration (H2,3), as shown in Fig. \ref{fig1}(d) and (e).

Upon increasing the applied magnetic field, the helical state degeneracy is lifted, and the structure transforms to the conical state, illustrated in Fig. \ref{fig1}(f), consisting of a continuous rotation of spins at an acute angle to the propagation vector, which is aligned to the applied field. In the field-parallel configuration, no magnetic scattering is observed, as displayed in Fig. \ref{fig1}(g), while in the field-perpendicular orientation, a single pair of conical magnetic satellites are observed, labelled C in Fig. \ref{fig1}(h). At higher applied magnetic field, the angle of the spins to the propagation vector is reduced, resulting in a lower intensity of the magnetic diffraction peaks, before a uniform magnetization texture is reached.

Close to 55 K and 22 mT, the skyrmion state is formed in a plane perpendicular to the applied magnetic field, as illustrated in Fig. \ref{fig1}(i). This image highlights the elongated tube-like structure exhibited by magnetic skyrmions in bulk materials \cite{milde_unwinding_2013}. In the field-parallel configuration, three pairs of magnetic scattering peaks are observed, a characteristic signature of the hexagonal skyrmion lattice, labelled S in Fig. \ref{fig1}(j). Due to rotational disorder of the skyrmion lattice, weak scattering can be observed in the field-perpendicular configuration, as shown in Fig. \ref{fig1}(k). This corresponds to the intensity at the top of the ring of scattering in (j). For further analysis in this study, we report the intensities of the H1, H2, H3, C and S magnetic peaks. These intensities were calculated by summing the total counts measured inside the red sector boxes specified in Fig. \ref{fig1}(d-k).

\section{AC Susceptibility Phase Diagrams}
\label{s:acsus}

\begin{figure}
\centering
\includegraphics[width=0.5\textwidth]{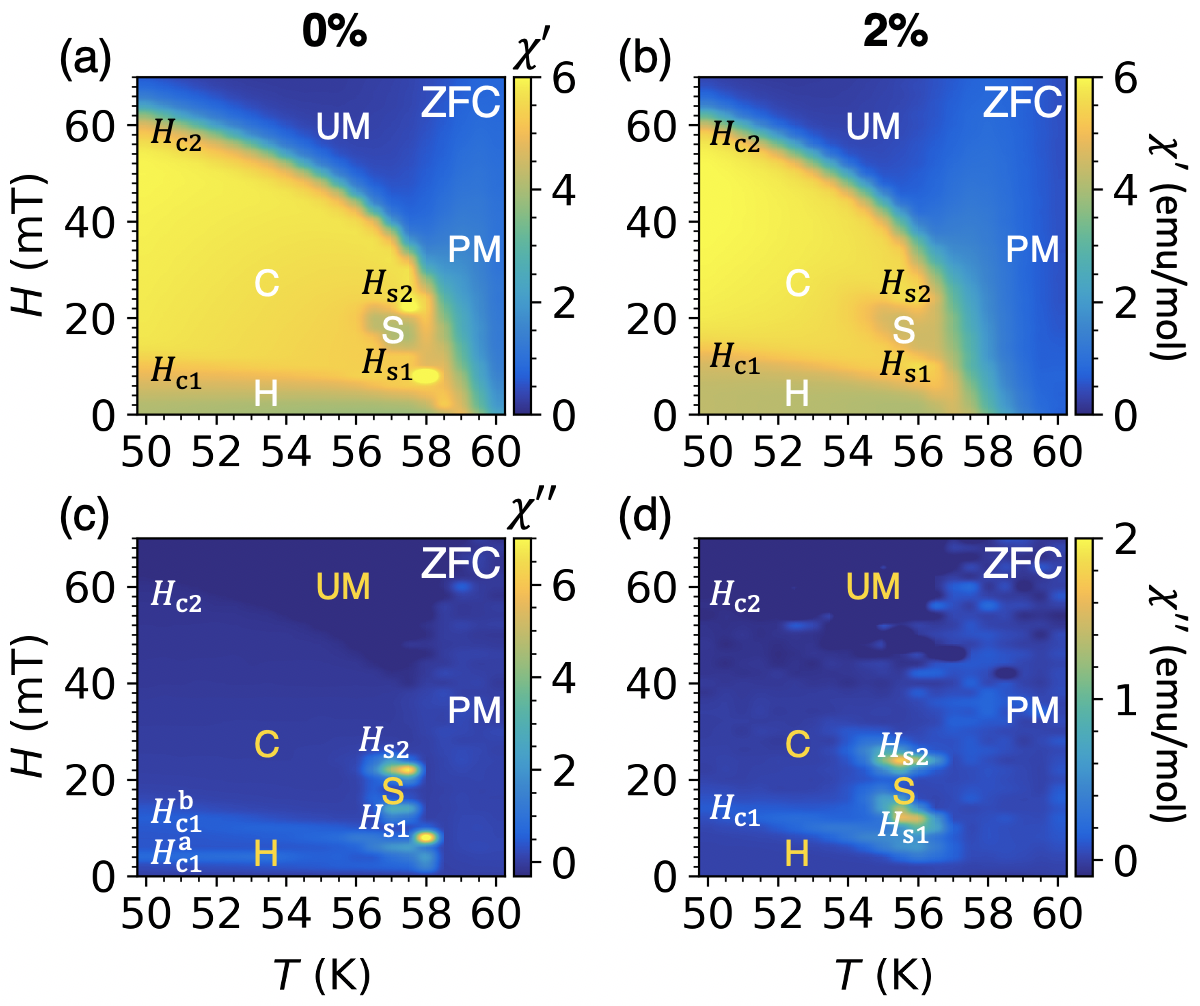}
\caption{(a-d), Magnetic phase diagrams, as determined by AC susceptibility measurements following the ZFC procedure, for the 0\% and 2\% Zn-substituted samples respectively, highlighting the helical (H), conical (C), skyrmion (S), uniform magnetization (UM), and paramagnetic (PM) states. The real component $\chi'$ is plotted in (a,b), and the imaginary component $\chi''$ in (c,d). Critical fields marking the phase boundaries between the H and C ($H_{\text{c1}}$), C and UM ($H_{\text{c2}}$), S and C ($H_{\text{s1,2}}$) states are labelled.}
\label{fig2}
\end{figure}

Magnetic phase diagrams are shown in Fig. \ref{fig2} for the pristine and 2\% Zn-substituted samples measured using AC susceptibility magnetometry. In Fig. \ref{fig2}(a) and (b), the colormap directly plots the real component of the AC susceptibility, $\chi'$, measured as a function of applied magnetic field after ZFC to temperatures between 50 and 60 K. At the low frequency limit, this quantity measures the local gradient of the magnetization with the applied field, and is useful for distinguishing the temperature and field extent of the magnetic phases. 

The boundaries between these magnetic phases are denoted by critical fields, $H_{\text{x}}$. The helical state can be identified by the reduced value of $\chi'$ close to 0 mT, and extending for increasing field up to $H_{\text{c}1}$, at the boundary with the conical state. In turn, the conical state exists up to $H_{\text{c}2}$, before transitioning to the field-polarized state, characterized by a reduction in $\chi'$. Close to $T_{\text{C}}$, the skyrmion state can be identified by the characteristic dip in AC susceptibility at $\sim$20 mT, a well-established signature common to skyrmion hosting bulk chiral magnets \cite{wilhelm_precursor_2011}. The lower and upper field boundaries of this phase are denoted $H_{\text{s}1}$ and $H_{\text{s}2}$ respectively. Looking at this real component, the phase diagrams for both the pristine and substituted sample show no dramatic differences besides the reduction of $T_\text{C}$ with increased Zn content.

At specific probe frequencies, dynamic processes may be excited, causing a phase difference between the oscillating magnetization and the drive field. This manifests as a signal in the imaginary component of the AC susceptibility, $\chi''$, indicating relaxation-induced energy losses, \cite{topping_ac_2018}. The dissipation is associated with thermodynamically irreversible dynamic processes such as domain wall motion, or, at a phase boundary, excitation between two competing magnetic phases \cite{balanda_ac_2013}. The $\chi''$ component of the AC susceptibility for both samples is plotted in Fig. \ref{fig2}(c) and (d). 

Second order phase transitions, characterized by the continuous transformation of one state into another, are not expected to exhibit a $\chi''$ signal. This can be seen by the lack of $\chi''$ peak at $H_{\text{c}2}$ in both Fig. \ref{fig2}(c) and (d), as the conical state continuously deforms to the uniform magnetization state. On the other hand, for first order phase transitions, characterized by coexistence of the two phases in the vicinity of the transition, a $\chi''$ signal can be expected: there is significant energy loss occurring as the oscillating magnetic field drives the boundaries between the two coexisting phase volumes \cite{halder_thermodynamic_2018}. Such an effect is observed at the edges of the skyrmion phase in Fig. \ref{fig2}(c) and (d), $H_{\text{s}1}$ and $H_{\text{s}2}$, where a large $\chi''$ signal is exhibited, indicating the annihilation and formation of skyrmions to and from the conical state \cite{bauer_history_2016}. Further $\chi''$ peaks seen at $H_{\text{c}1}$ indicate the helical to conical first order phase transition. 

The magnitude of $\chi''$ peaks associated with first order magnetic phase transitions can be modified by several factors \cite{balanda_ac_2013}. Firstly, $\chi''$ exhibits a frequency dependence, with the maximum indicating the resonant frequency of the excited dynamic processes \cite{bannenberg_magnetic_2016,qian_phase_2016}. Therefore, as we only measure at 10~Hz, a change in the size of the $\chi''$ peaks could be associated with a shift of this resonant frequency, or a broadening of the resonance due to the introduction of a range of relaxation timescales \cite{levatic_dissipation_2014}. Secondly, the peak can be reduced when phase coexistence in the first order transition is suppressed, as there can be no excitation between the two magnetic states, and therefore no energy losses \cite{halder_thermodynamic_2018}. Finally, when considering the decay of a metastable state, transitions to the ground state are energetically preferred. Thus, the oscillating field may only drive the transition in one direction, and no dissipative signal is expected \cite{halder_thermodynamic_2018}, despite the coexistence of the magnetic states.

\section{Temperature-Dependent AC Susceptibility Data}
\label{s:tempdep}

\begin{figure*}
\centering
\includegraphics[width=0.85\textwidth]{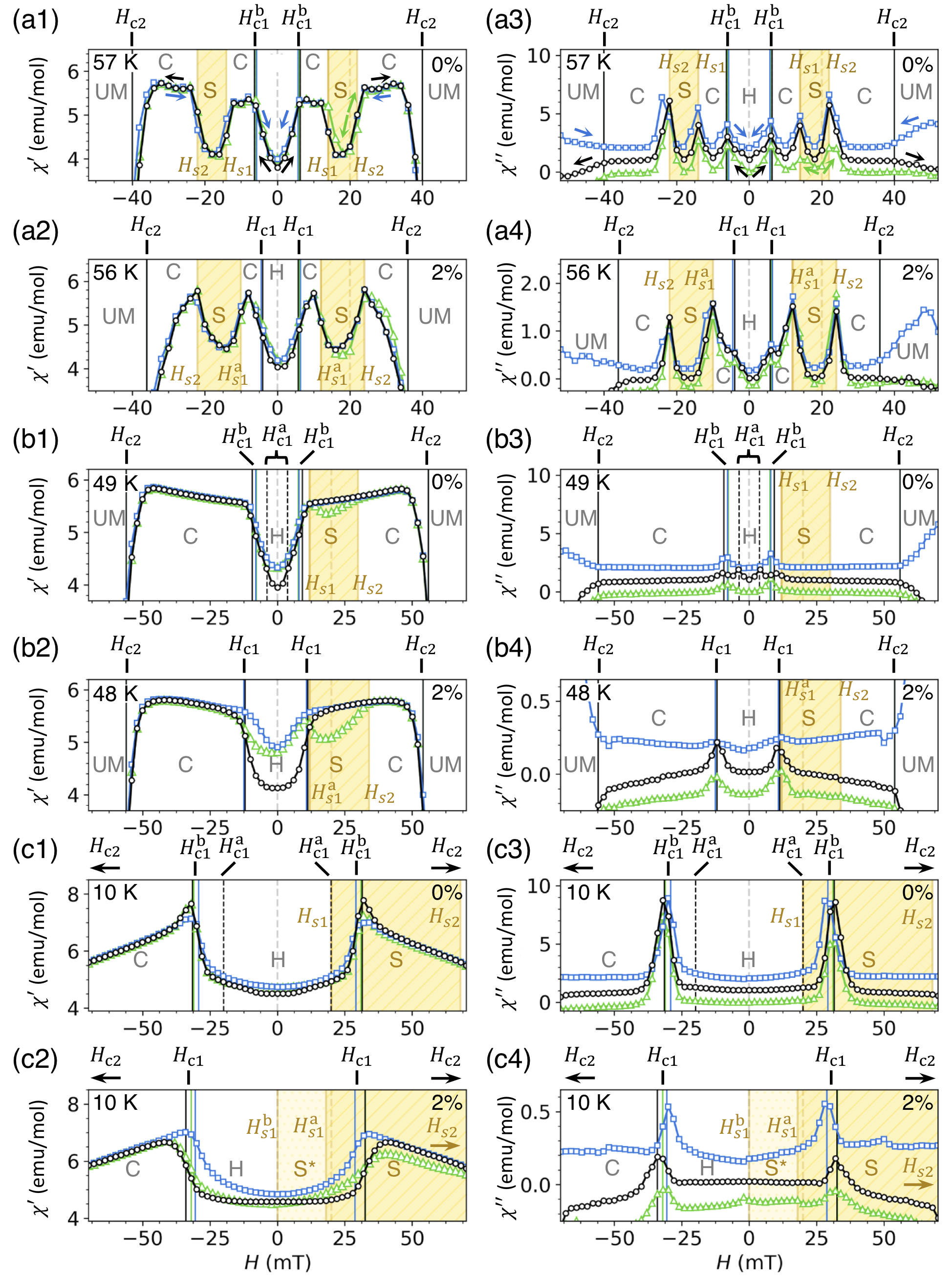}
\caption{(a)-(c)) The real, $\chi'$ (1 and 2), and imaginary, $\chi''$ (3 and 4), components of the AC susceptibility data measured as a function of magnetic field at selected temperatures for the pristine (1 and 3) and substituted (2 and 4) samples. Data acquired following the ZFC (black circles), HFC (blue squares) and FC (green triangles) measurement procedures is shown. The $\chi''$ data has been vertically offset to allow the features to be inspected as follows:  0\% ZFC: $+1.0$, 0\% FC: $+2.0$; 8\% HFC: $+0.15$, 8\% FC: $-0.15$. Coloured vertical lines indicate the fitted critical field values for each phase transition following each measurement procedure. The yellow filled regions indicate the extent of the skyrmion state after FC. The location of the helical (H), conical (C), skyrmion (S, S$^{\ast}$) and uniform magnetization (UM) states are labelled.}
\label{fig3a}
\end{figure*}

AC susceptibility data sets measured at a range of temperatures for both the pristine and substituted samples, and following the ZFC, HFC and FC measurements procedures, are shown in Fig. \ref{fig3a}. Critical field points at each phase boundary are indicated by the coloured vertical lines, and were determined by features in the data. The $H_{\text{c}2}$ points were estimated to be at the inflection point in the $\chi''$ data at high fields, most easily seen in Fig. \ref{fig3a}(a3). Below the equilibrium skyrmion region, no obvious feature exists for the upper skyrmion phase boundaries, $H_{\text{s}2}$, and the majority of the lower boundaries, $H_{\text{s}1}$. Therefore, these values were estimated by comparing features in the $\chi'$ data following the FC and ZFC/HFC procedures. For all the $H_{\text{c}1}$ points, and for $H_{\text{s}1}$ points below 40~K in the substituted sample, the field value was obtained by fitting a Gaussian peak with a linear background to the associated peak in the $\chi''$ data.

A summary of the fitted/estimate values of the critical fields, $H_{\text{x}}$, following each measurement procedure, are plotted as a function of temperature in Fig. \ref{fig3c}(a) and (b), forming an extended magnetic phase diagram for the pristine and substituted samples. In Fig. \ref{fig3c}(c) and (d), the fitted intensity of the $\chi''$ peaks (taken as the area under each peak) at $H_{\text{c}1}$ and $H_{\text{s}1}$ are plotted as a function of temperature for each sample. In the following analysis, we shall explore the detailed features of the AC susceptibility data at each temperature in Fig. \ref{fig3a}, while discussing the implications of these behaviours with reference to the summary panels in Fig. \ref{fig3c}.
 
\subsection{Equilibrium Skyrmion Phase}
\label{ss:equi}

We first consider the data in Fig. \ref{fig3a}(a), recorded at 57~K and 56~K for the pristine and substituted samples respectively. At these temperatures, the samples exhibit the equilibrium skyrmion phase, indicated by the yellow filled regions at positive and negative applied magnetic fields. The difference in the overall shape of the $\chi'$ data between the samples, shown in  Fig. \ref{fig3a}(a1, a2), can be attributed to the fact the measurements were not taken at the exact same temperature relative to $T_\text{C}$ ($T_\text{C}-T = 1.8$ and $1.3$ K for the pristine and substituted samples respectively). For each sample at this high temperature, there are no obvious differences in the data between the ZFC, HFC and FC measurement procedures: the fitted value of all critical fields are consistent with each other. 

However, comparing Fig. \ref{fig3a}(a3) and (a4), there is a significant difference in the magnitude of the $\chi''$ peaks in the substituted sample relative to the pristine sample. Such a compositional effect has been observed in Mn$_{1-x}$Fe$_x$Si, where the $\chi''$ signal around the skyrmion phase was reduced for $x>0.03$ \cite{bannenberg_magnetization_2018}, and it was argued that this indicated a decrease in the resonant frequency of the transition due to a slowing down of the dynamics by pinning effects. Therefore, following the arguments in Sec. \ref{s:acsus}, the reduction of the $\chi''$ signals in our substituted sample could either be caused by a broadening of the frequency dependence, or the introduction of metastable effects due to the additional pinning dynamics. The SANS data explored in later sections sheds some light on this issue, but thoroughly distinguishing between these scenarios requires future measurement of frequency dependent data.

\subsection{Helical-Conical Phase Transition}
\label{ss:heli-con}

\begin{figure}
\centering
\includegraphics[width=0.33\textwidth]{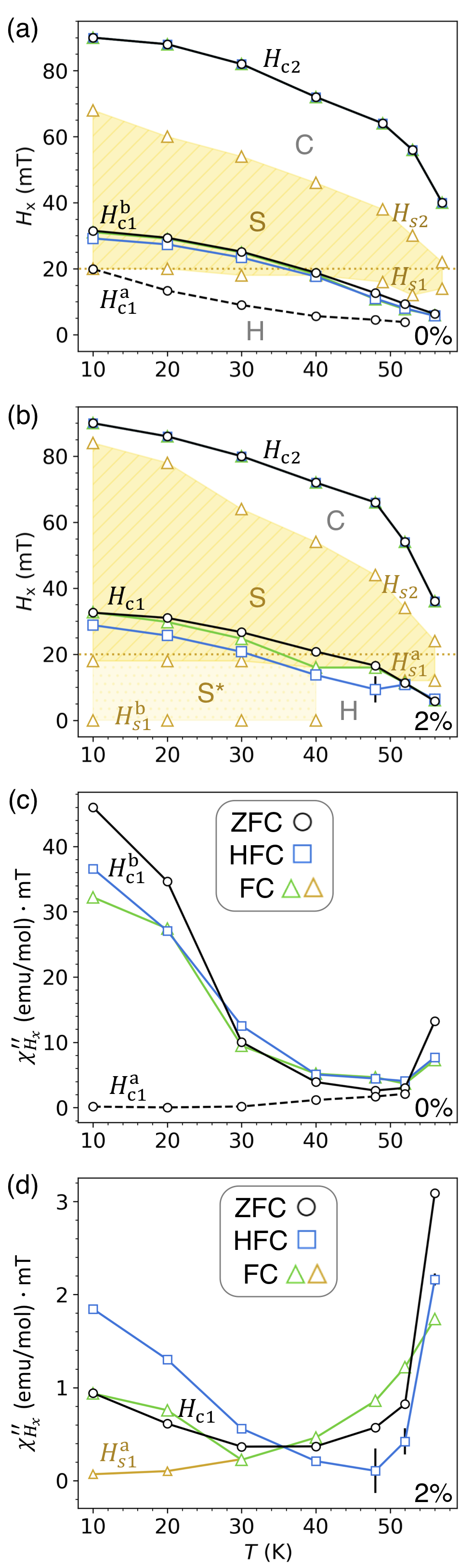}
\caption{(a,b), Low temperature magnetic phase diagrams for the pristine and substituted samples, formed by plotting the critical field values determined for each measurement procedure versus temperature, as labelled. Filled yellow regions designate the existence of the skyrmion state after FC. The location of the helical (H), conical (C), skyrmion (S, S$^{\ast}$) and uniform magnetization (UM) states are labelled. (c,d), The fitted intensity of the peak in the $\chi''$ data for the $H_{\text{c}1}$ and $H_{\text{s}1}$ critical fields after each measurement procedure for the pristine and substituted samples respectively.}
\label{fig3c}
\end{figure}

Next, we shall consider the behaviour of the helical-conical phase transition as a function of temperature. Turning first to the $\chi'$ data at 49 K in Fig. \ref{fig3a}(b1), there is a difference in the measured value of $\chi'$ around 0 mT following the ZFC and HFC procedures: the value of $\chi'$ it larger at 0 mT after HFC than it is after ZFC. This behaviour indicates that, when following the HFC procedure, the conical to helical phase transition results in a helical state different to the one formed upon ZFC -- perhaps with altered relative volumes of the H1 and H2,3 helical domains. This effect is more pronounced in the substituted sample, as can be seen in Fig. \ref{fig3a}(b2), suggesting that the Zn-substitution is hindering the conical to helical phase transition in some manner. However, for both samples, this low field divergence between the HFC and ZFC data is reduced at the lower temperature of 10 K, as shown in Fig. \ref{fig3a}(c1, c2). This is surprising, as one might expect the effects of pinning in the substituted sample to be more prevalent at lower temperatures, as there is less thermal energy available to enable depinning. 

In the $\chi''$ data, we see that at temperatures lower than the equilibrium skyrmion pocket, only peaks associated with helical-conical phase transition are seen. Unlike at high temperatures in Fig. \ref{fig3a}(a), at 49~K and below, we can see that there is a field offset in the peak position of each measurement procedure, indicating history-dependent behaviour in the $H_{\text{c}1}$ phase transition point. The fitted $H_{\text{c}1}$ values for both samples are summarized in Fig. \ref{fig3c}(a) and (b) respectively, highlighting that this hysteretic field offset becomes significant below 50~K, and is more pronounced in the substituted sample.

At 49~K and 48~K, it is evident that, in both samples, the magnitude of the $\chi''$ peaks are greatly reduced relative to the data measured at 57~K and 56~K. However, remarkably, looking at the data in Fig. \ref{fig3a}(c3, c4), we can see that the $\chi''$ peaks are actually larger in size at the lower temperature 10~K. This temperature dependence is made clear in the plot of the fitted $H_{\text{c}1}$ intensities in Fig. \ref{fig3c}(c) and (d). Comparing the data for the two samples, it is clear that this low temperature increase is far greater in the pristine sample, where the intensities of the $H_{\text{c}1}$ peaks surpass the measured intensities at 57~K. However, the substituted sample still displays a marked increase below 30~K in comparison to the values at 48~K. 

Once again, this could be due to a shift in the frequency dependence, suppression of phase coexistence, or metastable effects. However, considering frequency dependence, it is not possible for the $\chi''$ peak height to first reduce and then increase if the resonant frequency varies monotonically with temperature. Furthermore, one would expect that metastable effects would suppress the peak height more at lower temperatures, due to the reduction of available thermal energy. In comparison to other skyrmion materials, the different low temperature behaviour observed in the $\chi''$ data here in Cu$_2$OSeO$_3$ is significant, and will be explored and explained in the context of the SANS data in later sections.

\subsection{Metastable Skyrmions}
\label{ss:metastable}
Finally, the features associated with metastable skyrmions in the FC data will be examined. In Fig. \ref{fig3a}(b1) and (b2), both samples exhibit a drop in $\chi'$ around 20 mT after FC. This is indicative of the formation of metastable skyrmions at 48~K and 49~K, as displayed by the yellow hashed regions. The signal is far greater in the substituted sample, suggesting that the volume of metastable skyrmions which survived the cooling process was far greater in comparison to the pristine sample, as observed previously \cite{birch_increased_2019}. When considering the extent of this metastable skyrmion region, we can see that there is a lack of corresponding $\chi''$ signals in Fig. \ref{fig3a}(b3) and (b4). Following the argument mentioned previously, this can be expected for a first order phase transitions from a metastable state, where the oscillating probe field only drives the transition in one direction.

Nevertheless, we can estimate the skyrmion phase boundary using the $\chi'$ data. It is expected that as the skyrmions annihilate at higher fields, the value of $\chi'$ in the FC data will increase until it follows the ZFC values, as shown in Fig. \ref{fig3a}(b1, b2) and (c1, c2), indicating all skyrmions have decayed to the conical state. We use this assumption to estimate the $H_{\text{s}2}$ point for both samples at all temperatures, as summarized by the filled yellow region in Fig. \ref{fig3c}(a) and (b).

For decreasing field, determining the value of $H_{\text{s}1}$ using the $\chi'$ data is more complicated. If the skyrmions decay into the conical state before the helical state becomes energetically favoured at the $H_{\text{c}1}$ point, then we can expect the FC data to follow the HFC data to 0 mT. This is the scenario observed at higher temperatures in both samples, for example in Fig. \ref{fig3a}(b1) and (b2), suggesting that metastable skyrmions are annihilated before, or at, the helical-conical boundary. 

At lower temperatures, the FC data in the pristine sample follows the HFC data below 20 mT, suggesting that few, or possible no, metastable skyrmions survived the cooling process down to 10~K. A previous work looking at a pristine Cu$_2$OSeO$_3$ sample also noted a lack of metastable skyrmions surviving in this region with the field applied parallel to $[110]$ \cite{bannenberg_multiple_2019}. In contrast, for the substituted sample, below 20 mT the FC data follows closer to the ZFC data, as seen in Fig. \ref{fig3a}(c1, c2). This suggests that Zn substitution prevents the metastable skyrmions from decaying into the helical state during the FC process. Future SANS studies will be useful to clarify this behaviour.

Examination of the FC $\chi''$ data at 10~K in \ref{fig3a}(c3, c4) further supports this interpretation. In the substituted sample, at 30~K and below, two additional small peaks are exhibited. Since they appear only in the FC data, and only in the substituted sample, it is reasonable to assume that these are associated with the presence of metastable skyrmions. The first feature, labelled $H_{\text{s}1}^a$, appears just below the cooling field of 20 mT. The second feature, labelled $H_{\text{s}1}^b$, is a small peak just below 0 mT. We have distinguished this possible low field skyrmion region, S*, by the second filled yellow area in Fig.\ref{fig3a}(c4) and Fig.\ref{fig3c}(d). However, we argued previously that the annihilation of metastable skyrmions should exhibit no $\chi''$ signal, due to the lack of dissipation in metastable phase transitions. Therefore, further study will be required to fully understand the origin of these dynamic features, and whether they are truly associated with the presence of metastable skyrmions.

\section{Small Angle Neutron Scattering}
\label{s:sans}

\begin{figure}
\centering
\includegraphics[width=0.45\textwidth]{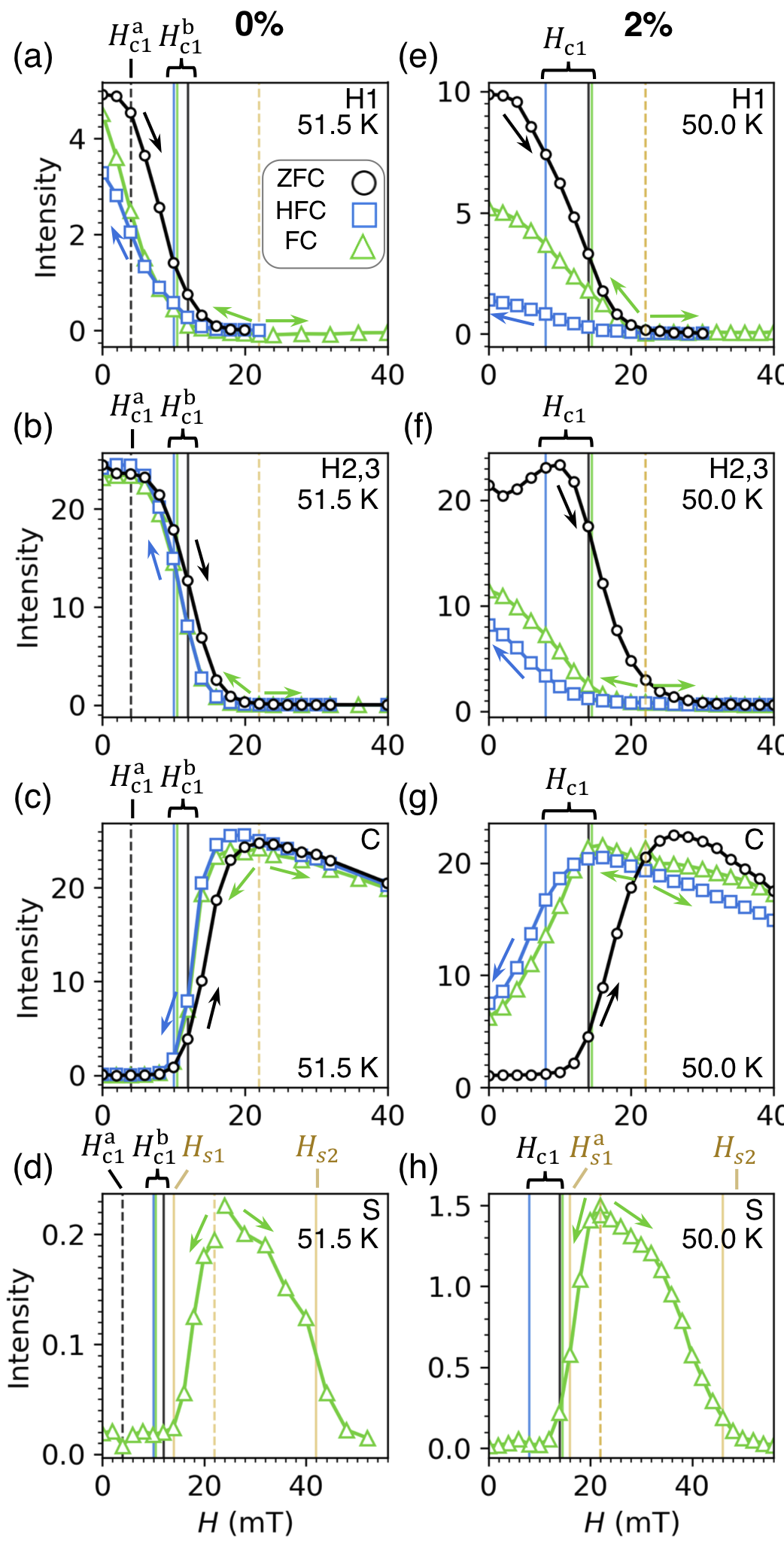}
\caption{The neutron scattering intensity of the H1, H2,3, C and S peaks measured in the pristine sample at 51.5~K (a-d) and the substituted sample at 50.0~K (e-h) as a function of applied magnetic field for the ZFC (black circles), HFC (blue squares) and FC (green triangles) measurement procedures. The intensities shown were determined by summing the total counts in the sector boxes as defined in Fig. \ref{fig1}. Vertical lines indicate critical fields determined by AC susceptibility measurements. Error bars are too small to be seen.}
\label{fig5}
\end{figure}

\begin{figure*}
\centering
\includegraphics[width=0.85\textwidth]{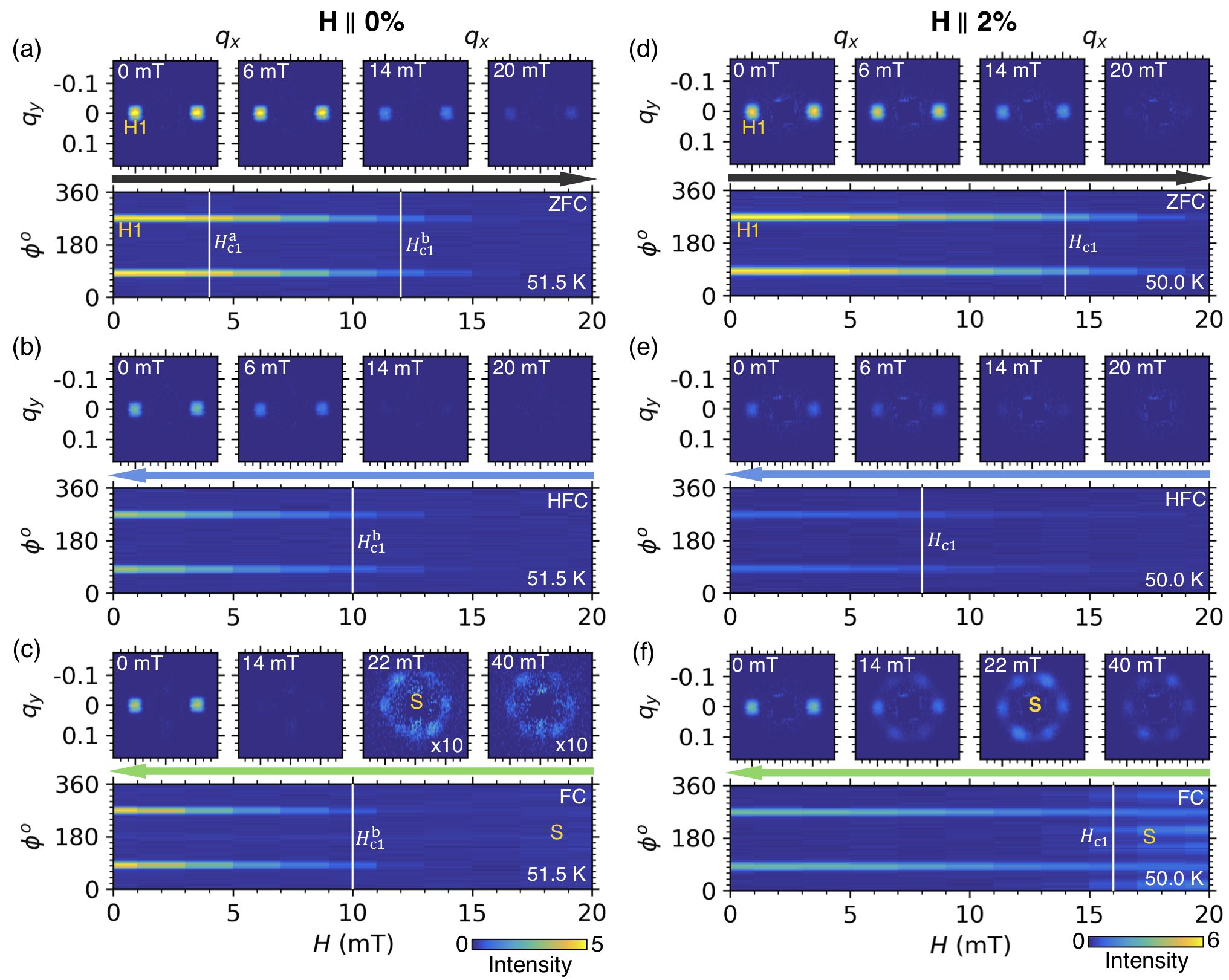}
\caption{SANS patterns measured following the ZFC, HFC and FC procedures in the field parallel configuration with the pristine sample at 51.5~K (a-c), and the substituted sample at 50.0~K (d-f). The upper panels display SANS patterns at selected fields, while the lower panels display the radially integrated intensity as a function of field and azimuthal angle, $\phi$, around the ring of scattering. Vertical lines indicate critical fields determined by AC susceptibility measurements. The units of $q_{x,y}$ are nm$^{-1}$. For the data presented in each sample configuration, the colorscale has been fixed to enable direct comparison between images.}
\label{fig6}
\end{figure*}

\begin{figure*}
\centering
\includegraphics[width=0.85\textwidth]{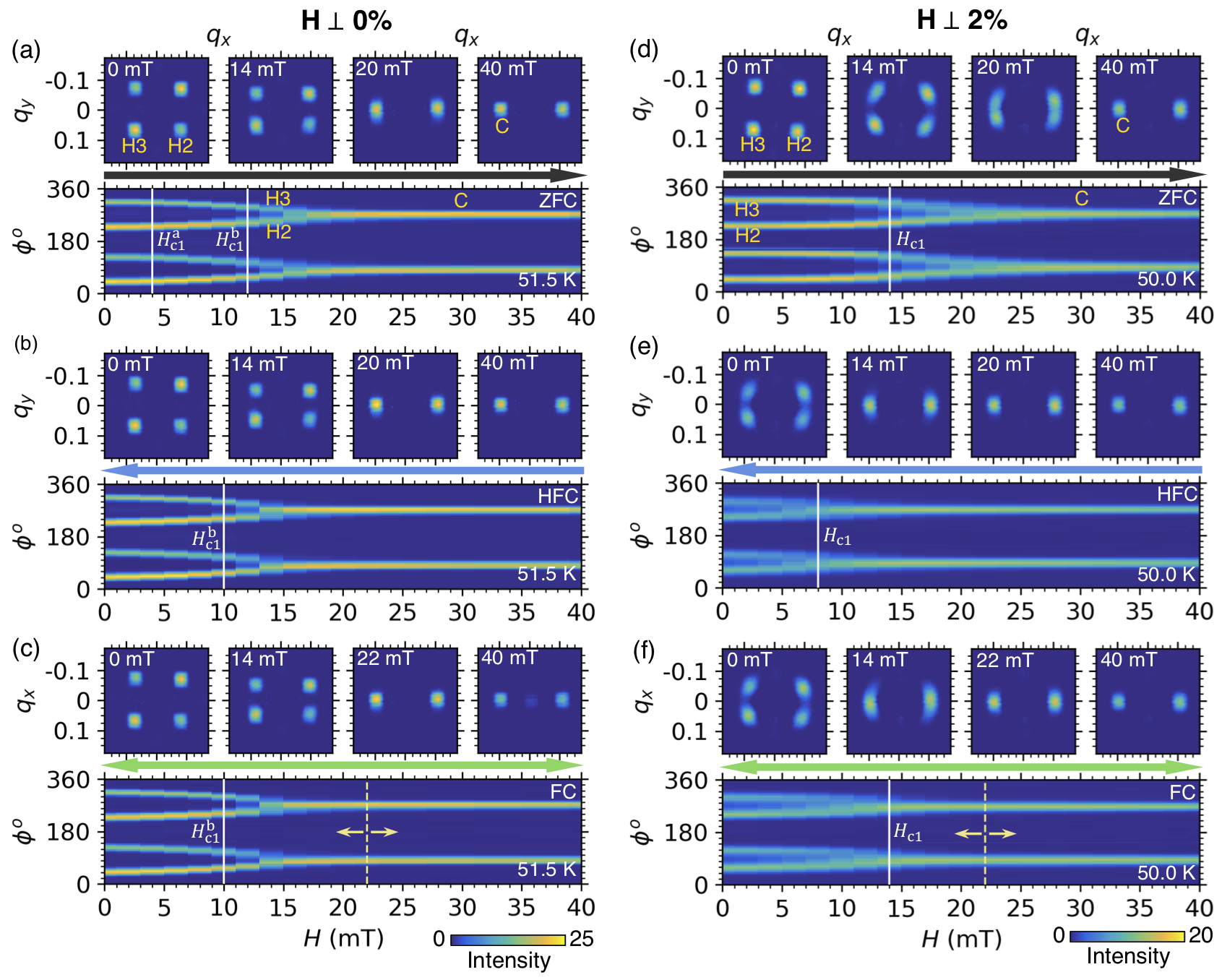}
\caption{SANS patterns measured following the ZFC, HFC and FC procedures in the field perpendicular configuration with the pristine sample at 51.5~K (a-c), and the substituted sample at 50.0~K (d-f). The upper panels display SANS patterns at selected fields, while the lower panels display the radially integrated intensity as a function of field and azimuthal angle, $\phi$, around the ring of scattering. Vertical lines indicate critical fields determined by AC susceptibility measurements. The units of $q_{x,y}$ are nm$^{-1}$. For the data presented in each sample configuration, the colorscale has been fixed to enable direct comparison between images.}
\label{fig7}
\end{figure*}

While the AC susceptibility measurements provide useful information, their interpretation is only able to provide a higher-level indication of the underlying associated phenomena. However, SANS measurements provide a more detailed picture of the microscopic behaviour, allowing the relative volumes and behaviour of each magnetic structure to be compared and contrasted when following each measurement procedure. We performed SANS measurements at both $\sim$50~K (Fig. \ref{fig5}, \ref{fig6}, \ref{fig7}) and 5~K, (Fig. \ref{fig8}, \ref{fig9}, \ref{fig10}), with the samples in both the field parallel and field perpendicular configurations. At both temperatures, and for both samples, the intensity of the H1, H2,3, C and S magnetic peaks are plotted as a function of applied magnetic field in Fig. \ref{fig5} and Fig. \ref{fig8}. These intensities were determined by summing the scattering in the sector boxes defined in Fig. \ref{fig1}(d-k) at each field point. The vertical lines in each figure show the critical fields determined by AC susceptibility measurements at the same temperatures, following the ZFC, HFC and FC procedures. 

Representative SANS patterns for these field scans are displayed in the upper panels of each subplot in Fig. \ref{fig6} and \ref{fig7}, and Fig. \ref{fig9}, \ref{fig10}. For each scattering pattern, the intensity over a $q$ range of 0.07 and 0.13 nm$^{-1}$ was summed radially at each azimthual angle $\phi$ around the centre of diffraction, as illustrated in Fig. \ref{fig1}(g). By reducing the dimensionality of the data in this way, the behaviour and evolution of the magnetic scattering peaks at each temperature can be examined as a function of the applied magnetic field in a single colour plot, as shown in the bottom panel of each subplot in Fig. \ref{fig6}, \ref{fig7} ($\sim$50~K), and Fig. \ref{fig9}, \ref{fig10} (5~K). In the following sections, we shall examine and interpret this SANS data while making reference to the features of the corresponding AC susceptibility data in Fig. \ref{fig3a} and Fig. \ref{fig3c}. 

\subsection{High Temperature SANS}
\label{ss:hitemp}

We first consider the SANS data measured at $\sim$ 50~K. The intensity of H1 in the pristine sample, plotted in Fig \ref{fig5}(a), exhibits a different low field behaviour for the ZFC, and HFC measurements. This can be seen clearly in the SANS patterns in Fig. \ref{fig6}(a,b), where the intensity of the H1 magnetic satellites at 0 mT is comparatively lower after HFC. This effect is enhanced in the substituted sample, showing an even greater difference between the ZFC and HFC measurements in Fig. \ref{fig5}(e), as illustrated by the SANS data in In Fig. \ref{fig6}(d,e). This suggests that during the transition from the conical state to the helical state with decreasing magnetic field, the H1 helical domains are formed at a lower volume fraction, with the Zn-substitution accentuating this effect. This fits with the observations gleaned from the $\chi'$ data around 0 mT in Fig. \ref{fig3a}(b), where the HFC value is greater compared to the ZFC value, which hinted at altered helical domain populations. This phenomena might be expected when considering the orientation of the conical and helical domains in Fig. \ref{fig1}(c) and (f): the conical domains must rotate by 45 degrees to reorient to the H2,3 helical domains, which might require less energy than the 90 degree rotation required to reorient to the H1 helical domain, and it appears that pinning in the substituted sample accentuates this effect. 

In both Fig. \ref{fig5}(d) and (h) the skyrmion scattering intensity for each sample is displayed, exhibiting the broad range of field over which metastable skyrmions exist. The $H_{\text{s1}}$ and $H_{\text{s2}}$ values estimated from the features seen in the $\chi''$ data in Fig. \ref{fig3a}, align well with the field extent of the skyrmion scattering in both Fig. \ref{fig5}(d) and (h), and illustrate that, at this temperature, the metastable skyrmions decay before the conical-helical phase boundary. The skyrmion peaks can be seen clearly in Fig. \ref{fig6}(c) and (f). The greater relative intensity of the S satellites in the substituted sample confirms the formation of a larger population of metastable skyrmions, as was indicated by the AC susceptibility measurements.

In both samples, the H1 intensity at 0 mT is higher after FC in comparison to HFC. Given that the primary difference between these two measurement procedures is the formation and subsequent annihilation of metastable skyrmions, this suggests that the skyrmion state preferentially annihilates to H1 helical domains upon decreasing magnetic fields. This finding fits with the previously proposed skyrmion lattice to helical state decay process, whereby skyrmion tubes are zipped together through the motion of Bloch points \cite{milde_unwinding_2013}, forming helical domains. In this process, formation of H1 domains would likely be preferred over the H2,3 domains, because they are oriented in the same plane as the skyrmion lattice, as can be seen by comparing Fig. \ref{fig1}(c) and (f). 

The intensities of H2,3 and C for the pristine sample, plotted in Fig. \ref{fig5}(b) and (c), shows only minor differences between the three procedures, with a slight field offset in the observed behaviour. The SANS patterns for the ZFC procedure in Fig. \ref{fig7}(a) demonstrate that, at this temperature, the helical satellites rotate to their position in the conical state as a function of applied magnetic field. This indicates that the propagation vectors of the H2 and H3 helical domains themselves continuously rotate during the phase transition, maintaining long-range magnetic order throughout, with no coexistence of helical and conical states. This behaviour is replicated for the reverse conical to helical phase transition, as shown in Fig. \ref{fig7}(b,c). These processes, occurring as a continuous rotation of the magnetic structure, appear more like a second order, rather than first order, phase transition. The observed suppression of helical and conical phase coexistence provides an explanation of the reduction of the $H_\text{c1}$ $\chi''$ peak at these temperatures in the AC susceptibility data in Fig. \ref{fig3c}, as argued in Sec. \ref{s:acsus}.

On the other hand in the substituted sample, the intensity of the H2,3 peaks and C peaks, plotted in \ref{fig5}(f) and (g), show significant differences between the ZFC, HFC and FC measurements. The ZFC SANS patterns in Fig. \ref{fig5}(d) depict a gradual rotation of the helical domains to the conical state, similar to that of the pristine sample, but occurring at a higher applied magnetic field. However, at 0 mT in the HFC and FC measurements, in Fig. \ref{fig5}(e) and (f), we see that the rotation of the conical state back to the helical state is incomplete: the magnetic structures remain partially pinned along the applied field direction. These observations strongly agree with the features seen in the AC susceptibility data at these temperatures in Fig \ref{fig3a}(b): the substituted sample displayed a larger discrepancy in the value of $\chi'$ between the ZFC and HFC data around 0 mT, which hinted at an altered helical state after HFC. The AC data also showed an even greater suppression of the $\chi''$ signal at $H_{\text{s1}}$, and this is explained by the observed metastable effects in the HFC SANS data. In comparison to the pristine sample, this hysteretic behaviour indicates that the presence of Zn in the substituted sample introduces a pinning effect which hinders the reorientation of the helical and conical states.

\subsection{Low Temperature Behaviour}

\begin{figure}
\centering
\includegraphics[width=0.45\textwidth]{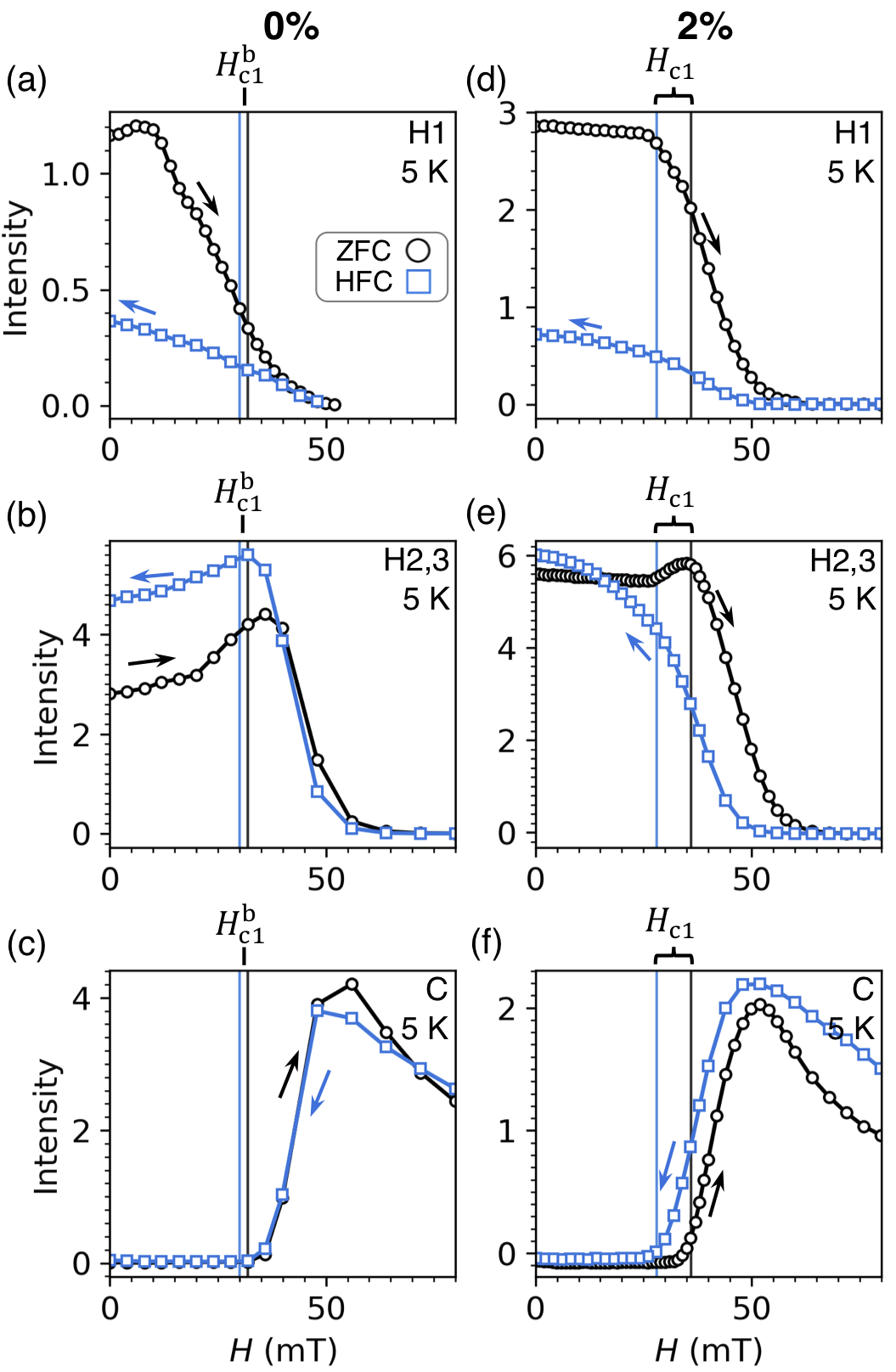}
\caption{The neutron scattering intensity of the H1, H2,3 and C peaks measured in the pristine sample (a-c) and the substituted sample (d-f) at 5~K as a function of applied magnetic field for the ZFC (black circles) and HFC (blue squares) measurement procedures. The intensities shown were determined by summing the total counts in the sector boxes as defined in \ref{fig1}. Vertical lines indicate critical fields determined by AC susceptibility measurements. Error bars are too small to be seen.}
\label{fig8}
\end{figure}

\begin{figure*}
\centering
\includegraphics[width=0.835\textwidth]{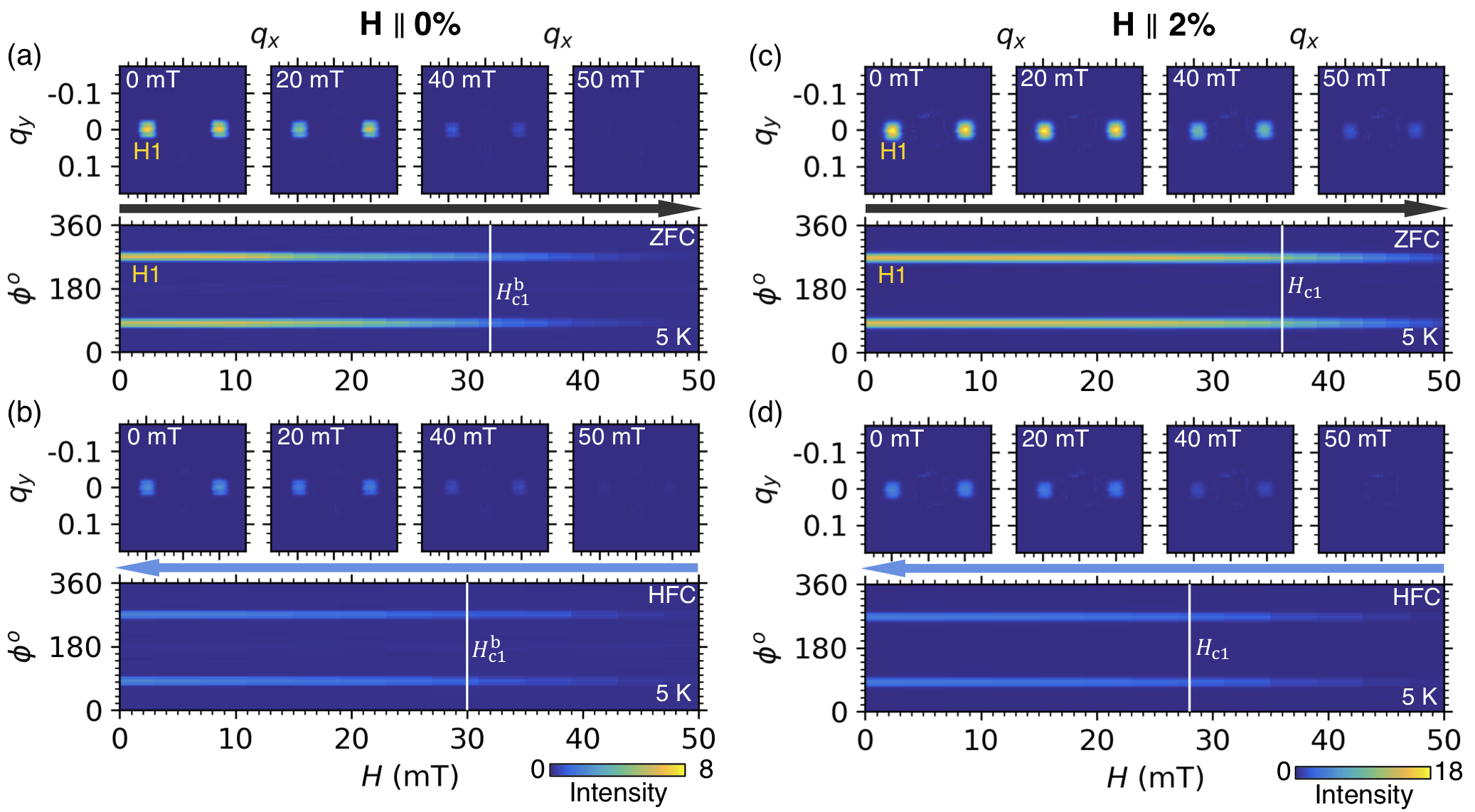}
\caption{SANS patterns measured following the ZFC and HFC procedures in the field parallel configuration with the pristine sample (a, b), and the substituted sample (c, d) at 5~K. The upper panels display SANS patterns at selected fields, while the lower panels display the radially integrated intensity as a function of field and azimuthal angle, $\phi$, around the ring of scattering. Vertical lines indicate critical fields determined by AC susceptibility measurements. The units of $q_{x,y}$ are nm$^{-1}$. For the data presented in each sample configuration, the colorscale has been fixed to enable direct comparison between images.}
\label{fig9}
\end{figure*}

\begin{figure*}
\centering
\includegraphics[width=0.835\textwidth]{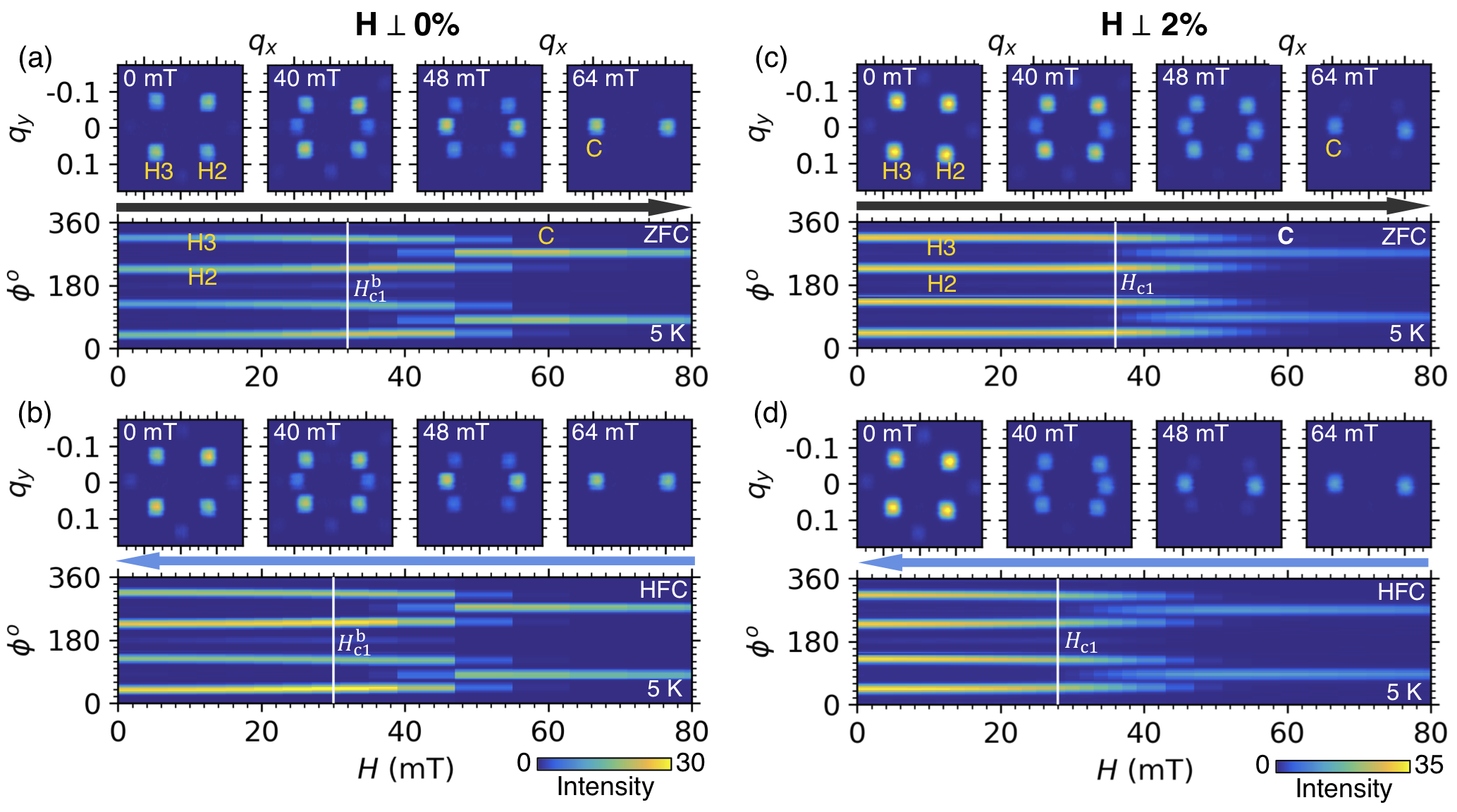}
\caption{SANS patterns measured following the ZFC and HFC procedures in the field perpendicular configuration with the pristine sample (a, b), and the substituted sample (c, d) at 5~K. The upper panels display SANS patterns at selected fields, while the lower panels display the radially integrated intensity as a function of field and azimuthal angle, $\phi$, around the ring of scattering. Vertical lines indicate critical fields determined by AC susceptibility measurements. The units of $q_{x,y}$ are nm$^{-1}$. For the data presented in each sample configuration, the colorscale has been fixed to enable direct comparison between images.}
\label{fig10}
\end{figure*}

We turn now to the SANS measurements performed at 5~K, shown in Fig. \ref{fig8}, \ref{fig9} and \ref{fig10}. Looking at the ZFC data for the pristine sample in Fig. \ref{fig8}(a), the H1 peak gradually decreases in intensity at higher fields. Simultaneously, the intensity of the H2,3 peaks initially increases, before decreasing after the critical field of $H_{\text{c}1}^{b}$, where the conical state becomes favoured, as shown in Fig. \ref{fig8}(b). In contrast, in the ZFC data of the substituted sample, shown in Fig. \ref{fig8}(d) and (e), the H1 and H2,3 peak intensities are almost constant until just before $H_{\text{c}1}$ at $\sim$ 30~mT. Here H1 begins to decrease, while the H2,3 intensity shows a small increase approaching $H_{\text{c}1}$, before decreasing. 

This could indicate that, in the pristine sample, the population of H1 helical domains steadily transforms to H2,3 helical domains as a function of field, as the degeneracy of the helical domains is lifted under an increasing applied magnetic field, and the H2,3 helical domains are energetically favoured before $H_{\text{c}1}^{b}$. On the other hand, in the substituted sample, this processes is largely suppressed. The observed behaviour may explain the presence of the additional $H_{\text{c}1}^{a}$ $\chi''$ peaks observed during ZFC in the pristine sample in Fig. \ref{fig3a}, but not in the substituted sample. Another explanation of this behaviour could be a distortion of the helical state with increasing field \cite{izyumov_modulated_1984}, which would explain the drop in H1 intensity, but not the increase in H2,3 intensity. 

Turning now to the HFC measurements for the pristine sample, in comparison to the 51.5 K data, we see an further reduction in the H1 intensity at 0 mT relative to the ZFC data, as seen in Fig. \ref{fig8}(a) and Fig. \ref{fig9}(b). Considering Fig. \ref{fig8}(d) and Fig. \ref{fig9}(d), we can see that the difference in HFC behaviour between the pristine and substituted samples is less pronounced than it was at $\sim$50~K. This suggests that at lower temperatures, the reduction in H1 domain population after HFC is largely due to energetic considerations, and pinning from the chemical substitution has less effect.

The intensity of the H2,3 helical domain peaks measured in both the pristine and substituted samples are shown in Fig. \ref{fig8}(b) and (e). Looking at both the H2,3 and C peak intensities, there is a the dramatic field offset in the observed trends for the substituted sample, agreeing with the significant hysteresis displayed by the $\chi''$ peak position in AC data relative to the pristine sample in Fig \ref{fig3a}. In the pristine sample, the H2,3 intensity at 0 mT is higher after HFC than in the ZFC process, again suggesting that H2,3 helical domains are favoured when the conical state transforms to the helical state with decreasing field. 

In the 50~K measurements of the substituted sample, the H2,3 satellites showed a large reduction in intensity at 0 mT for the HFC measurements, caused by pinning of the magnetic domains as they rotated from the conical to the helical state. However, notably, at this lower temperature, H2,3 intensity is higher at 0 mT after HFC in comparison to the ZFC data, exhibiting similar low temperature behaviour to the pristine sample. This is surprizing, as one might expect the effects of pinning to be greater at lower temperatures.

Fig. \ref{fig10} reveals the details of the altered low temperature behaviour. In Fig. \ref{fig10}(a) and (b), we see the helical to conical phase transition in the field-perpendicular configuration. At 51.5 K, the phase transition was characterized by a continuous rotation of the magnetic structures as a function of applied field, and phase coexistence was suppressed. In contrast, at 5~K, the simultaneous detection of both helical and conical peaks in one SANS pattern indicates phase coexistence, as expected for a first order phase transition. 

The altered low temperature behaviour is more clearly evident in the 2\% sample, as shown in Fig. \ref{fig10}(c) and (d).  After HFC at 50~K, the Zn-substitution hindered the continuous rotation of the conical structure to the helical state. However, at 5~K we see that at 0 mT the helical domains fully reorient to the [100] and [010] directions after HFC. As in the pristine sample, there is phase coexistence of the helical and conical states for both ZFC and HFC processes. This implies that the pinning which hindered the domain reorientation at higher temperature does not affect the phase transition in the same manner at low temperature.

\section{The Role of Cubic Anisotropy}
\label{s:cubicaniso}

\begin{figure}
\centering
\includegraphics[width=0.5\textwidth]{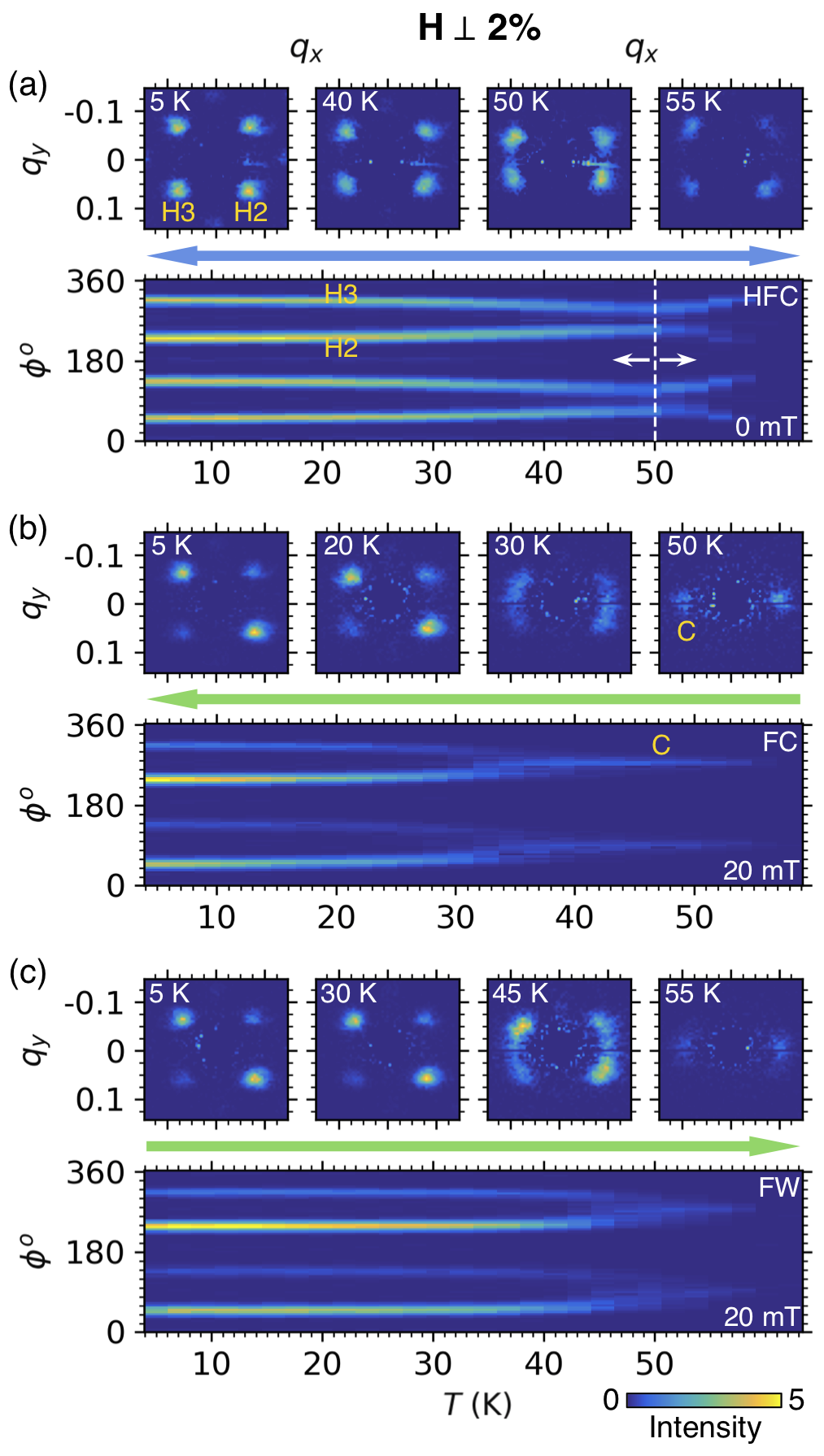}
\caption{SANS data measured as a function of temperature after high-field cooling (HFC) to 0 mT (a), while field cooling (FC) to 5~K (b), and field warming to 58~K (c). The upper panels display SANS patterns at selected temperatures, while the lower panels display the radially integrated intensity as a function of temperature and azimuthal angle, $\phi$, around the ring of scattering. The units of $q_{x,y}$ are nm$^{-1}$.}
\label{fig11}
\end{figure}

To investigate and understand this behaviour further, we performed a set of temperature dependent SANS measurements on the substituted sample in the field perpendicular orientation. For the first measurement, the sample was HFC at 200~mT to 50~K and then the magnetic field was decreased to 0~mT, forming the pinned helical state, as seen previously in Fig. \ref{fig7}(e). From this point, SANS measurements were performed as the sample was cooled down to 5~K, and, after resetting the magnetic state upon a subsequent cooling procedure, as the sample was heated from 50~K to 58~K. The resulting SANS data is shown in  Fig. \ref{fig11}(a). For increasing temperature data, the partially pinned helical state gradually rotates and orients along the $\langle100\rangle$ axes at higher temperatures, before disappearing at $T_\text{C}$. This can be expected, as at higher temperatures there is more energy available to overcome the pinning energy. However, we can see that upon decreasing the temperature from 50~K to 5~K, the helices start to reorient to the $\langle100\rangle$ axes below 40~K, despite the reduction in available thermal energy. This suggests that another contribution must be providing the energy to overcome the pinning at low temperatures.

In these skyrmion-hosting helimagnets, the orientation of the helical domains is determined by the cubic anisotropy. As discussed in the introduction, recent studies have shown that the anisotropy constant in Cu$_2$OSeO$_3$ greatly increases in magnitude below 40~K \cite{halder_thermodynamic_2018}. This provides an explanation for the observed behaviour in this study: the cubic anisotropy becomes strong enough at low temperature to overcome the pinning of the conical-helical reorientation in both the pristine and substituted sample. From the previously reported AC susceptibility measurements we have surveyed and referenced, it appears that the larger low temperature $\chi''$ signal at the helical-conical phase boundary is unique to Cu$_2$OSeO$_3$, suggesting the cubic anisotropy term is responsible for the emergence of helical-conical phase coexistence at low temperatures.

There are other noticeable effects of the increased cubic anisotropy strength in Cu$_2$OSeO$_3$. In the archetypal skyrmion material MnSi, it has been observed that the $H_{\text{c1}}$ value does not change dramatically as a function of temperature \cite{bannenberg_magnetization_2018}. Since the value of $H_{\text{c1}}$ in an indicator of the strength of the cubic anisotropy, this illustrates that the anisotropy does not significantly change as function of temperature. In comparable chemically doped materials such as Fe$_{1-x}$Co$_x$Si and Mn$_{1-x}$Fe$_x$Si, the measured value of $H_{\text{c1}}$ greatly increases with decreasing temperature, however this is suggested to be due to the pinning of the helical-conical phase transition \cite{munzer_skyrmion_2010,bannenberg_extended_2016,bannenberg_magnetic_2016,bannenberg_magnetization_2018}. This is concluded by noting that, upon decreasing the field from the conical state, such as after HFC or FC, the pinning effects are strong enough to fully prevent reorientation to the helical state, even close to $T_\text{C}$ \cite{bauer_history_2016}. This is similar to the behaviour we saw at 50~K in the Zn-substituted sample in this study.

In contrast, the behaviour of $H_{\text{c1}}$ as a function of temperature in both the pristine and substituted Cu$_2$OSeO$_3$ samples in this study are remarkably similar. This is highlighted in Fig \ref{fig3c}(a) and (b): both samples display an increase in $H_{\text{c1}}$ at lower temperatures. Although the substituted sample does exhibit an offset in $H_{\text{c1}}$ between ZFC and HFC measurements, these are small compared to the effect seen in other chemically substituted materials. This limited history-dependent behaviour in both samples indicates that the dominant determinant of $H_{\text{c1}}$ value in Cu$_2$OSeO$_3$ is the strong, temperature dependent cubic anisotropy, which, at low temperature, is able to largely overcome the effects of pinning introduced by the non-magnetic Zn dopant.

A final consideration in the context of the role of cubic anisotropy is the behaviour of metastable skyrmions. In previous studies, when FC through the skyrmion pocket in the materials such as MnSi, Fe$_{1-x}$Co$_x$Si and Mn$_{1-x}$Fe$_x$Si, the skyrmion state coexists with the conical state down to base temperature, and the helical state is not observed \cite{bauer_history_2016,bannenberg_magnetic_2016}. On the other hand, in Cu$_2$OSeO$_3$, due to the increase of $H_{\text{c1}}$ at low temperatures, the helical phase manifests below 35~K when FC the sample through the skyrmion region at 20 mT, as seen by the transition of the C to H2,3 magnetic satellites in Fig. \ref{fig8}(b). A similar transition is seen upon field warming (FW) in Fig. \ref{fig11}(b), but at a higher temperature. As speculated when considering the FC $\chi'$ data in Fig. \ref{fig3a}, it is likely that the emergence of the helical state while FC results in the loss of metastable skyrmion population. Such losses might be limited by avoiding the helical phase during the FC process: by cooling at 20~mT to 40~K, increasing the field to 40~mT, and then continuing the cooling process.

\section{Conclusions}
\label{s:conclusions}
A combination of detailed AC susceptibility magnetometry and small angle neutron scattering was utilized to study the phase transitions between, and relative volumes of, the helical, conical and skyrmion states in both pristine and Zn-substituted Cu$_2$OSeO$_3$. Comparison of the data between the samples, and three distinct measurement protocols, resulting in a number of observations and conclusions.

Measurements in both samples have demonstrated that upon HFC, the H1 helical domain volume fraction at 0mT is reduced after HFC in comparison to ZFC at both high and low temperature, with the effect greater in the Zn-substituted sample. This can be understood when considering the 90 degree rotation required to transform the conical state to the H1 helical orientation, which likely imposes a larger energy barrier in comparison to the 45 degree rotation required to form the H2,3 helical domains. In contrast, the larger H1 intensity at 0~mT after the FC protocol indicates that the metastable skyrmion state preferentially decays to H1 helical domains, lending support to the previously proposed skyrmion-to-helical decay mechanism: the zipping together of skyrmion tubes via the motion of magnetic Bloch points.

In the pristine sample, an additional low field peak in the ZFC $\chi''$ data suggested a further phase transition, which was identified to be H1 helical domains transforming to H2,3 domains as the helical domain degeneracy is lifted by the applied field before the helical-to-conical phase boundary at $H_{\text{c1}}$. The lack of this peak in the substituted sample suggests that this reorientation is effectively prevented by the introduction of pinning.

At 51.5~K, the pristine sample SANS data revealed that the helical-conical phase transition, for both ZFC and HFC procedures, is characterized by a continuous rotation of the H2,3 helical domains towards the conical state as a function of the applied field, and suppression of phase coexistence. This was accompanied by a dramatic reduction in the corresponding $\chi''$ dissipation peak, indicative of suppression of phase coexistence in a first order phase transition. In the substituted sample at 50.0~K, the magnetic texture did not fully reorient to the $\langle100\rangle$ helical domain axes upon decreasing field to 0 mT after HFC, suggesting they were partially pinned along the magnetic field direction. This behaviour appears to be consistent with that seen in other substituted helimagnets, where the disorder introduces a similar pinning effect.

In contrast, at 5~K, the SANS data illustrated that the same phase transition displayed coexistence of the helical and conical phases. The corresponding peaks in the $\chi''$ data greatly increasing in magnitude at lower temperature, as expected for a first order transition exhibiting phase coexistence. Remarkably, the pinning effects which dominated the magnetic reorientations in the substituted sample at 50~K, appear to be overcome at this lower temperature, despite the reduction of available thermal energy. In comparison to other archetypal helimagnets, this behaviour appears to be unique to Cu$_2$OSeO$_3$. We attribute this low temperature behaviour to the large, temperature dependent cubic anisotropy energy present in Cu$_2$OSeO$_3$: at low temperature, the anisotropy becomes strong enough to overcome the pinning energy in the absence of thermal activation. The relatively moderate history-dependence of the helical-conical phase boundary in both the pristine and substituted samples demonstrates the dominance of this aniosotropic energy contribution. 

Overall, these results highlight further unique behaviour exhibited by the magnetic phase transitions in Cu$_2$OSeO$_3$ due to its comparatively large, temperature dependent cubic anisotropy. Comparisons between pristine and substituted samples reveal the role of disorder in the slowing down and pinning of the helimagnetic phase transition dynamics, highlighting that these may be effectively overcome by other energy contributions. Consideration and further study of these effects will be crucial when utilizing chemical substitution, or doping, to manipulate the delicate energy balance for exploitation in future applications of skyrmion materials.

\begin{acknowledgments}
This work was supported by the UK Skyrmion Project EPSRC Programme Grant (EP/N032128/1). Experiments at ILL neutron source were supported by experiment time awarded under proposal no. 5-42-488. Experiments at the ISIS Pulsed Neutron and Muon Source were supported by a beamtime allocation from the Science and Technology Facilities Council, proposal number 1920655. M. N. Wilson acknowledges the support of the Natural Sciences and Engineering Research Council of Canada (NSERC).
\end{acknowledgments}

\end{document}